%% file: LLM_judge.tex
\documentclass[sigconf,natbib=true,anonymous=false]{acmart}
\usepackage{subcaption}
\usepackage{graphicx}
\usepackage{multirow}
\usepackage{makecell}
\usepackage{enumitem}
\usepackage[dvipsnames]{xcolor}
\usepackage[most]{tcolorbox}
\usepackage{cleveref} 
\setlist{leftmargin=10pt}

\usepackage{capt-of}% or \usepackage{caption}
\usepackage{booktabs}
\usepackage{varwidth}
\newcommand{\sys}[1]{\textsf{#1}}
\newcommand{\qwen}{\sys{Qwen}}
\newcommand{\gemma}{\sys{Gemma}}
\newcommand{\llama}{\sys{Llama}}
\newcommand{\mistral}{\sys{Mistral}}
\newcommand{\qwenfull}{\sys{Qwen-3-8B}}
\newcommand{\gemmafull}{\sys{Gemma-3-4B}}
\newcommand{\llamafull}{\sys{Llama-3.2-3B}}
\newcommand{\mistralfull}{\sys{Mistral-7B}}
\newcommand{\gemini}{\sys{Gemini-2.5-Flash}}
\newcommand{\passive}{\textsf{PSV}}
\newcommand{\activ}{\textsf{ACV}} % JMM: \active is already a keyword
\newcommand{\summary}{\textsf{SUM}}
\newcommand{\expansion}{\textsf{EXP}}
\newcommand{\sem}{\textsf{SEM}}
\newcommand{\lex}{\textsf{LEX}}
\newcommand{\qry}{\textsf{QRY}}
\newcommand{\myparagraph}[1]{\paragraph*{\hspace*{-\parindent}\normalsize\bf#1}}

\AtBeginDocument{%
  \providecommand\BibTeX{{%
    \normalfont B\kern-0.5em{\scshape i\kern-0.25em b}\kern-0.8em\TeX}}}

\copyrightyear{2026}
\acmYear{2026}
\setcopyright{cc}
\setcctype{by-nc-nd}
\acmConference[SIGIR '26]{Proceedings of the 49th International ACM SIGIR Conference on Research and Development in Information Retrieval}{July 20--24, 2026}{Melbourne, VIC, Australia}
\acmBooktitle{Proceedings of the 49th International ACM SIGIR Conference on Research and Development in Information Retrieval (SIGIR '26), July 20--24, 2026, Melbourne, VIC, Australia}
\acmDOI{10.1145/3805712.3809905}
\acmISBN{979-8-4007-2599-9/2026/07}

\begin{document}

% \title{LLMs Can’t Replace Human Assessors: Understanding Overrating and the Limits of Semantic Comprehension}
% \title{LLMs Can't Replace Human Assessors: \\Overrating and the Limits of Semantic Relevance}
% Joel says: Do not remove the square brackets, or a newline will be added to the page headers where the title is 
\title{When LLM Judges Inflate Scores: Exploring Overrating in Relevance Assessment}

\author{Chuting Yu}
\orcid{}
\email{v.yu@uq.edu.au}
\affiliation{%
  \institution{The University of Queensland}
  \city{Brisbane}
  \country{Australia}
}

\author{Hang Li}
\email{hang.li@uq.edu.au}
\affiliation{%
  \institution{The University of Queensland}
  \city{Brisbane}
  \country{Australia}
}

\author{Guido Zuccon}
\email{g.zuccon@uq.edu.au}
\affiliation{%
    \institution{The University of Queensland}
    \city{Brisbane}
    \country{Australia}
}

\author{Joel Mackenzie}
\email{joel.mackenzie@uq.edu.au}
\affiliation{%
  \institution{The University of Queensland}
  \city{Brisbane}
  \country{Australia}
}

\author{Teerapong Leelanupab}
\email{t.leelanupab@uq.edu.au}
\affiliation{%
  \institution{The University of Queensland}
  \city{Brisbane}
  \country{Australia}
}
\renewcommand{\shortauthors}{Yu et al.}

\begin{abstract}
Human relevance assessment is time-consuming and cognitively intensive, limiting the scalability of Information Retrieval evaluation. This has led to growing interest in using large language models (LLMs) as proxies for human judges. However, it remains an open question whether LLM-based relevance judgments are reliable, stable, and rigorous enough to match humans for relevance assessment. In this work, we conduct a study of \textit{overrating behavior} in LLM-based relevance judgments across model backbones, evaluation paradigms (pointwise and pairwise), and passage modification strategies. We show that models consistently assign inflated relevance scores --- often with high confidence --- to passages that do not genuinely satisfy the underlying information need, revealing a system-wide bias rather than random fluctuations in judgment. Furthermore, controlled experiments show that LLM-based relevance judgments can be highly sensitive to passage length and surface-level lexical cues. These results raise concerns about the usage of LLMs as drop-in replacements for human relevance assessors, and highlight the urgent need for careful diagnostic evaluation frameworks when applying LLMs for relevance assessments. Our code and results are publicly available.\footnote{\textcolor{RoyalBlue}{\url{https://github.com/chutingyu/Exploring-Overrating}}}

% Human relevance assessment is time-consuming and label-intensive, limiting the scalability of Information Retrieval (IR) evaluation. This has led to growing interests in using large language models (LLMs) as proxies for human judges. However, it remains an open question whether LLM-based relevance judgments are reliable, stable, and rigorous enough to match human assessments. In this work, we conduct a systematic study of overrating behavior in LLM-based relevance judgments across different model backbones, evaluation paradigms (pointwise and pairwise), model sizes, where models assign inflated relevance scores to passages that do not genuinely satisfy the underlying information need. Further examine whether inflated relevance scores are accompanied by inflated model confidence(token-level logits). Our findings reveal that overrating does not happen randomly but is a system-wide deficit. Furthermore, LLM-based relevance judgments can be highly sensitive to context and length even under controlled conditions. These results raise concerns about the usage of LLMs as drop-in replacements for human relevance assessors. Our results highlight the urgent need for careful diagnostic evaluation frameworks when applying LLMs for relevance assessments.

\end{abstract}

\begin{CCSXML}
<ccs2012>
   <concept>
       <concept_id>10002951.10003317.10003359.10003361</concept_id>
       <concept_desc>Information systems~Relevance assessment</concept_desc>
       <concept_significance>500</concept_significance>
       </concept>
 </ccs2012>
\end{CCSXML}

\ccsdesc[500]{Information systems~Relevance assessment}
% keywords
% \vspace{-0.2cm}
\keywords{LLMs, Overrating Bias, Relevance Judgment, Relevance Cues}

\maketitle

\input{1_introduction}

\input{2_overrating}
\input{3_relevance_cues}
\input{4_conclusion}

\input{5_future_works}

\begin{acks}
We thank the anonymous referees for their insightful comments and suggestions. This work was supported by the UQ HERA initiative and by a Google Research Scholar
grant.
\end{acks}

\balance % ensures reference columns are balanced 
\renewcommand{\bibsep}{1pt} % JMM: Space out the bib entries a bit
\bibliographystyle{ACM-Reference-Format}
\bibliography{references}

% \appendix









\end{document}

%% file: 1_introduction.tex
\section{Introduction and Background}

Building reliable relevance annotations for evaluation has long been a central challenge in Information Retrieval (IR). Cranfield-style collections {\cite{Cranfield}}, such as TREC DL~\cite{craswell2020overview,craswell2021overview}, rely on trained assessors to annotate query-document pairs, costing substantial time and effort~\cite{soboroff2025DontUseLLMs}, while crowdsourced annotations often suffer from limited reliability \cite{guo2023EvaluatingLargeLanguage, rahmaniJudgingJudgesCollection2025, tadele}. Large language models (LLMs) offer a potential alternative by enabling relevance judgments to be generated rapidly and at scale.

Recent studies report that LLM-based relevance judgments can closely align with human assessments. \citet{Thomas:2024:Large-La} employed real searcher feedback to select LLM–prompt configurations aligned with user preferences, demonstrating performance comparable to human assessors in identifying strong systems and difficult queries. These findings were reproduced and extended by \citet{upadhyay2024largescalestudyrelevanceassessments}, who further released the open-source UMBRELA framework~\cite{upadhyay2024umbrela}. ~\citet{Upadhyay:2025:A-Large-} showed that LLM-generated judgments can replace traditional human labels at the {\emph{run level}} in the TREC 2024 RAG Track, with limited benefit from additional human-in-the-loop intervention.

While LLMs have shown to be effective annotation tools, their potential shortcomings as relevance assessors remains insufficiently understood. LLM-based judgments raise several issues. For example, \citet{Alaofi:2024:LLMs-Can} show that LLMs are vulnerable to keyword stuffing, systematically overrating non-relevant passages that contain query terms by assigning inflated relevance labels. Inflated relevance labels, as compared to humans, have been reported in prior work~\cite{Upadhyay:2025:A-Large-,meng2025QueryPerformancePrediction}. This behavior also demonstrates how strong {\emph{aggregate}} performance can obscure judgment-level errors, which can directly impact downstream evaluation tasks. Moreover, using LLMs as both a retrieval method and as an evaluation (labeling) method introduces circularity, confounding true system improvements with LLM-specific preferences~\cite{Clarke:2025:LLM-base}. These issues highlight the potential risks when adopting LLMs as relevance assessors.

Since IR systems are measured over the set of relevance labels for a given set of topics, these labels serve as an upper bound on measurable effectiveness. When relevance judgments are generated by LLMs, this upper bound is implicitly shaped by the models’ own judgment behaviors. LLM-based relevance assessments tend to exhibit label inflation (e.g., assigning a document a higher relevance label than that assigned by a human assessor), which may distort evaluation signals and mask genuine differences in retrieval quality. However, it remains unclear how widespread this behavior is in different model families and evaluation settings.

In this work, we conduct a systematic analysis of overrating behavior in LLM-based relevance judgments across two TREC datasets (DL2019 and DL2020), and four open-source LLMs ({\llamafull}, {\gemmafull}, {\mistralfull}, and {\qwenfull}), on both pointwise and pairwise evaluation paradigms. We analyze overrating at both the label level, and via token-level confidence, to examine whether it reflects random fluctuations or a systematic model bias. To examine the role of semantic understanding, we further conduct controlled passage-rewriting experiments, varying passage length and syntactic voice\footnote{i.e., the grammatical relationship between a verb and its subject, determining whether the subject performs (active) or receives (passive) the action.} while preserving meaning, and experiment with varying the content of non-relevant passages to isolate the effects of lexical overlap and surface relevance cues.

%% file: 2_overrating.tex
\section{Overrating Behavior in LLM Judges}

We first analyze overrating in LLM-based relevance assessment in terms of both the labels, and the confidence of the assessment.

% Although score inflation in LLM-based relevance judgments has been observed in several studies~\cite{Alaofi:2024:LLMs-Can,Upadhyay:2024:A-Large-,meng2025QueryPerformancePrediction}, it remains unclear how widespread this behavior is in different models, model sizes, and evaluation settings. In this paper, we investigate whether LLMs exhibit overrating tendencies at both the score and confidence levels. 

\myparagraph{Label Overrating.}
Overrating in LLM-based relevance judgments occurs when LLMs assign quality valuations that exceed the ground truth established by human assessors.
In pointwise evaluation (binary and graded), overrating manifests through an upward shift in the distribution of LLM-assigned labels relative to human labels. 
%For binary labels (0/1), overrating occurs when non-relevant documents (human label 0) are incorrectly labeled as relevant (model label 1). In graded labeling schemes (0--3), overrating appears as upward grade shifts, where passages receive higher relevance levels from LLMs than those assigned by human assessors.
%
In pairwise evaluation, overrating is implicit due to the comparative nature of the task. Assuming that graded relevance labels imply a strict human preference (e.g., $2 > 0$),
two failure modes emerge: {\emph{lack of discrimination}}, where the model predicts a tie despite a clear human preference, and {\emph{preference hallucination}}, where a lower-grade passage is preferred over a higher-grade one.

% \begin{itemize}
% \item {Pointwise:}
% Overrating in pointwise evaluation (binary and graded) is explicitly quantitative. It appears as numeric score inflation, where the distribution of LLM-assigned scores systematically shifts toward the upper end of the scale compared to human labels.
% \begin{itemize}
% \item \textit{Binary Scoring (0/1):} Overrating is observed when the model incorrectly labeling non-relevant documents (Human Label~0) as relevant (Model Label 1). 
% \item \textit{Graded Relevance Scoring (0--3):} Overrating appears as numeric score inflation. Observed as upward score shifts, where the model assigns higher relevance grades than human annotators for the same passages.
% \end{itemize}

% \item {Pairwise:} Conversely, in pairwise settings, overrating is implicit due to the comparative nature of the task (e.g., A > B, or Tie). We identify two modes:
% \begin{itemize}
%     \item \textit{Lack of Discrimination (Tie):} The model declares a ''Tie'' when humans perceive a clear quality gap, effectively elevating the inferior passage to the level of the superior one.
%     \item \textit{Preference Hallucination (Wrong Wins):} A more severe form where the model explicitly prefers a lower grade (e.g., ''Unacceptable'') over a higher one (e.g., ''Best''). 
% \end{itemize}
% \end{itemize}

\myparagraph{Model Confidence and Stability.}
\label{sec-conf}
Beyond label inflation, we analyze the token-level confidence on the LLM's predicted label or preference. % We define confidence overrating as high confidence despite incorrect predictions in relevance judgments. %We define confidence overrating as cases where subjective certainty exceeds objective accuracy. 
%
% In the pointwise setting, we determine the prediction of {\emph{confidently incorrect}} as a model that assigns near-deterministic probabilities (i.e., softmax probabilities approaching 1.0) to incorrect label tokens, which are measured by the normalized logits on the generation of the label.
In the pointwise setting, we determine {\emph{confidently incorrect}} prediction as cases where the model assigns near-deterministic probabilities (i.e., softmax probabilities approaching 1.0) to incorrect label tokens, which are measured by the normalized logits on the generation of the label.
In the pairwise setting, instability manifests where the LLM  yields varying preferences under positional swapping (e.g., $A > B$ vs.\ $B > A$) while still assigning high confidence to the assessments.

\subsection{Experimental Setup}
\label{sub:exp_setup}
We conduct experiments on the TREC Deep Learning (DL) Track passage ranking tasks from 2019-2020~\cite{craswell2020overview,craswell2021overview}, using human~relevance \mbox{labels (0--3)} as ground truth. We evaluate four open-weight LLMs: {\llamafull}~\cite{grattafiori2024llama3herdmodels}, {\gemmafull}~\cite{gemma_2025}, {\mistralfull}~\cite{jiang2023mistral7b}, and {\qwenfull}~\cite{yang2025qwen3technicalreport} under both pointwise and pairwise settings. 

In the pointwise setting, we consider both binary relevance (0 ``Non-relevant'', 1 ``Relevant'') and graded relevance (0 ``Non-relevant'', 1 ``Related'', 2 ``Highly relevant'', 3 ``Perfectly relevant''). We apply variations of the UMBRELA prompt~\cite{upadhyay2024umbrela} for all judgment tasks. %Graded judgments are binarized according to the TREC standard for this task.\footnote{Not relevant: $0$ and $1$; Relevant: $\geq 2$. See: \url{https://trec.nist.gov/data/deep2019.html}} 
Each configuration is evaluated in terms of label overrating and model confidence.
%In addition, we conduct a controlled analysis within the {\qwen} model family, varying model size from \textsf{1.7B} to \textsf{32B} parameters to isolate the effect of model scale on overrating behavior.

% This is more appropriate place to say you vary model size of Qwen here, better than in introduction

\myparagraph{Pointwise Evaluation.} For pointwise judgments, we measure label inflation through two metrics: (1) Overrate Ratio, the proportion of the instances where the LLM assigns a relevance label exceeding the human label; and (2) Mean Bias, the average signed difference between LLM-assigned labels and human relevance labels (LLM minus Human). We additionally report Cohen's $\kappa$ to analyze agreement with human judgments.
%(3) Score Distribution Shift (LLM vs.\ Human), the overall relevance score distribution compared to human annotations.
These metrics collectively quantify both the magnitude and distributional patterns of label inflation exhibited by different LLMs. 
%
%To measure confidence, we compute the average token-level confidence (see Section~\ref{sec-conf}), and aggregate it into True Positive (TP), True Negative (TN), False Positive (FP), and False Negative (FN) bins based on the binarized labels. To construct the confusion matrix, graded judgments are binarized according to the TREC standard, allowing human and LLM judgment pairs to be categorized into the four classes above.\footnote{Not relevant: $0$ and $1$; Relevant: $\geq 2$. See: \url{https://trec.nist.gov/data/deep2019.html}} 
%and the Confidence Gap (FP$-$TP). 
To understand model confidence, we binarized both the human and LLM judgments according to the TREC standard.\footnote{Not relevant: $0$ and $1$; Relevant: $\geq 2$. See: \url{https://trec.nist.gov/data/deep2019.html}} Then, we aggregate the token-level confidence according to the True Positive (TP), True Negative (TN), False Positive (FP), and False Negative (FN) classes when comparing the human and LLM judgment pairs. Our intuition is that models may be less confident for incorrect (FP or FN) judgments.

\myparagraph{Pairwise Evaluation.}
For pairwise judgment, passages are grouped per query into three categories: {\emph{Best}} (top‐graded relevant), {\emph{Acceptable}} (relevant, but not the highest grade), and {\emph{Unacceptable}} (non‐relevant)~\cite{Arabzadeh:2025:Benchmar}. We sampled 20 pairs per query from each of the following settings: {\emph{Best~vs.~Acceptable}}; {\emph{Acceptable~vs.~Unacceptable}}; and {\emph{Best~vs.~Unacceptable.}} We treat the higher‐class passage as the ground‐truth winner, and pairs are judged in {\emph{both}} presentation orders. We force the LLM to prefer one of the two documents, rather than allowing an explicit tie option. As such, we define a {\emph{tie}} as the case where a contradiction arises between the \textit{two presentation orderings}, meaning that the judgments cannot be linearized. This indicates that the model cannot reliably distinguish the relevance difference between the passages. Therefore, we interpret such cases as the model treating the two passages as equally relevant.

Pairwise overrating behavior is measured using {\emph{Pairwise Accuracy}} (Acc), the overall proportion of correctly ordered pairs (including ties), and {\emph{Pairwise Accuracy without Ties}} (Acc (No Ties)), which excludes tied predictions and evaluates only non-tied comparisons. We also compute the {\emph{Tie Rate}}, capturing failures to discriminate between passages of unequal quality, and the {\emph{Wrong Wins Rate}}, measuring cases where the lower-grade passage is explicitly preferred over the higher-grade one in {\emph{both}} presentations.
We report the average token-level confidence on the set of correctly-ordered preferences and the set of incorrectly-ordered preferences.

\subsection{Results: Pointwise Judgment}
\begin{table*}[t]
\centering
\resizebox{0.9\textwidth}{!}{%
\input{tables/overrate-by-dataset}
}
\vspace{4pt}
\caption{
Pointwise overrating and underrating metrics on DL2019 and DL2020 under binary and graded relevance labels. Average Confidence is computed under the graded relevance label setting.
}
\label{tab:overrate-by-dataset}
% \vspace{-2pt}
\end{table*}

Table~\ref{tab:overrate-by-dataset} demonstrates the pervasive nature of overrating across all models and datasets from our results. We consistently observe that the overrating ratio is substantially higher than the underrating ratio, indicating LLMs rarely assign labels lower than human annotators. Since the two settings use different relevance scales~(Section~\ref{sub:exp_setup}), we binarized graded labels based on TREC standard to enable a fair comparison; the observed overrating ratio becomes lower than that obtained from direct binary evaluation. This suggests that a substantial portion of overrating errors in the graded setting occurs within the relevant or highly relevant categories. Cohen’s $\kappa$ is consistently lower in the graded setting than in the binary setting, which suggests that LLM judgments are less consistent with human annotations in finer-grained relevance, although agreement is typically lower in graded scenarios due to the additional freedom afforded by having more relevance levels.

% Table~\ref{tab:overrate-by-dataset} demonstrates the pervasive nature of overrating across all models and datasets from our results. In the binary setting, all models are substantially more conservative than in the graded setting, yielding markedly lower overrate rates. For instance, {\llama}’s overrate increases from 18.57\% (binary) to 51.11\% (graded). This indicates that when restricted to a relevant-irrelevant decision boundary, models apply a stringent relevance criterion. Under the finer-grained graded scale, LLM judges more often promote relevance levels, demonstrating substantial overrating behavior. This shift is also reflected in agreement with human labels: Cohen’s $\kappa$ is consistently lower in the graded setting than in the binary setting.

Examining the label transition matrices in the graded setting also reveals that the overrating behavior is dominated by soft errors: non-relevant passages are rarely assigned the highest relevance score, but are frequently misclassified as marginally or moderately relevant. This systematic reluctance to assign the non-relevant label inflates relevance estimates under graded evaluation. 

We determine whether model confidence (token-level logits) can serve as a reliable proxy for judgment quality in pointwise settings by analyzing the average confidence scores across TP, TN, FP, and FN predictions. Ideally, a reliable model should exhibit high confidence for correct predictions and noticeably lower confidence for errors. Our results indicate a profound disconnect between correctness and confidence.
Intuitively, TP decisions should carry noticeably higher confidence than FP decisions. However, our results reveal that the confidence gap is negligible or even inverted.
We observe a universal tendency towards extreme overconfidence across all models and tasks. In binary scoring, every model exhibits over 95\% confidence not only on TP and TN but also on FP and FN; we observe a similar pattern of high confidence in graded setting as showed in Table~\ref{tab:overrate-by-dataset}. This phenomenon reveals that, irrespective of the relevance of the passage, the model is almost always ``absolutely confident'' of its decision.

\subsection{Results: Pairwise Judgment}

\myparagraph{Trend 1: Lack of Discrimination.}
% \textit{Trend 1: Lack of Discrimination ("Tie").} 
As shown in Table~\ref{tab:pair_acc}, the ``Tie Rate'' is substantial for all models, ranging from approximately 24\% to over 45\%. This indicates order-induced inconsistency (preference flips under swapping), not consistent errors where the model repeats the same wrong choice. Furthermore, we investigated how judgment varies with the different graded pairs, as, intuitively, the {\emph{Best vs Unacceptable}} pairs should be the easiest to discriminate. However, we observed that all four LLMs tested consistently overrate ``Unacceptable'' passages, treating them as qualitatively equivalent to the ``Best'' passages in nearly {\emph{one-third}} of cases. Moreover, the tie rate increases systematically as the discrimination task becomes more difficult (e.g., {\emph{Best vs Acceptable}}), reflecting the models’ growing inability to distinguish between relevance grades on such higher difficulty tasks.

\myparagraph{Trend 2: Indecision vs. Hallucination.}
% \textit{Trend 2: The Nature of Overrating: Indecision vs. Hallucination.}
The most revealing commonality across all models is the source of error. By comparing accuracy and accuracy (no ties) in Table~\ref{tab:pair_acc}, a consistent behavioral pattern emerges: LLMs are accurate when they express a decisive preference, while overrating primarily arises from their tendency to assign equivocal judgments rather than making clear distinctions. Put simply, when ties are excluded, all models exhibit strong reliability. {\qwen} achieves an impressive accuracy (no ties) of 93\% on DL2020, and even the lowest performing model exceeds 85\%. These findings suggest that LLMs do not typically hallucinate that a bad passage is better than a good one (as demonstrated by the rather low ``Wrong Wins'' rate); rather, they fail to perceive the quality gap between the good and bad passages, defaulting to a Tie.

The confidence difference between correct and incorrect judgments is negligible in the pairwise scenario. On DL2019, {\gemma}'s average confidence drops only marginally from 0.979 on correct judgments (Corr.) to 0.964 on incorrect ones (Incorr.); {\llama} shows an even smaller gap. In other words, LLM confidence is virtually indistinguishable, regardless of judgment correctness.

\begin{table}[t]
\centering
\resizebox{0.46\textwidth}{!}{%
\input{tables/pair_acc}%
}
\vspace{4pt}
\caption{
Pairwise preference-judgment results on DL2019 and DL2020. Average confidence is computed on no-tie cases only for correct and incorrect preferences.
}
\label{tab:pair_acc}
% \vspace{-0.2cm}
\end{table}

Analyzing the model confidence on the tied pairs showed that on around 75\% of such cases, model confidence exceeded 0.95 on {\emph{both presentation orderings}} even though the preferences were contradictory, indicating a tendency for the LLMs to express overconfidence even when their judgments are logically inconsistent. Overall, these results demonstrate that using confidence as a signal for judgment reliability is not useful, and is highly sensitive to positional changes.

% Table~\ref{tab:pair_acc} shows the results of the pairwise judgment experiment. Overall, these results reveal universal trends across all evaluated models.%, regardless of their architecture or size, which are discussed in further detail below.

%% file: tables/overrate-by-dataset.tex
% \begin{tabular}{llcccccccccc}
% \toprule
% % \textbf{Dataset
% & \multirow{2}{*}{\textbf{Model}} &
% \multicolumn{2}{c}{\textbf{Overrated (\%)}} &
% \multicolumn{2}{c}{\textbf{Mean Bias}} &
% \multicolumn{2}{c}{\textbf{Cohen's $\kappa$}} &
% \multicolumn{4}{c}{\textbf{Avg Confidence}} \\

% \cmidrule(lr){3-4}
% \cmidrule(lr){5-6}
% \cmidrule(lr){7-8}
% \cmidrule(lr){9-12}

% & &
% \textbf{Bin.} & \textbf{Grad.} &
% \textbf{Bin.} & \textbf{Grad.} &
% \textbf{Bin.} & \textbf{Grad.} &
% \textbf{TP} & \textbf{FP} & \textbf{TN} & \textbf{FN} \\
% \midrule

% \multirow{4}{*}{\rotatebox[origin=c]{90}{\textbf{DL19}}}
% & \gemma    & 37.81  & 61.84  & 0.341  & 0.736 & 0.246 & 0.107 & 0.987 & 0.989 & 0.989 & 0.977 \\
% & \llama    & 18.82  & 49.72  & 0.077  & 0.564 & 0.304 & 0.163 & 0.902 & 0.881 & 0.958 & 0.914 \\
% & \mistral  & 21.53  & 54.23  & 0.141  & 0.557 & 0.369 & 0.150 & 0.939 & 0.932 & 0.951 & 0.904 \\
% & \qwen     & 20.62  & 47.02  & 0.145  & 0.459 & 0.421 & 0.215 & 0.977 & 0.976 & 0.991 & 0.986 \\
% \midrule

% \multirow{4}{*}{\rotatebox[origin=c]{90}{\textbf{DL20}}}
% & \gemma    & 44.78 & 66.51 & 0.431 & 0.895 & 0.163 & 0.088 & 0.983 & 0.979 & 0.986 & 0.981 \\
% & \llama    & 18.57 & 51.11 & 0.124 & 0.691 & 0.265 & 0.149 & 0.908 & 0.870 & 0.920 & 0.871 \\
% & \mistral  & 22.32 & 55.08 & 0.180 & 0.646 & 0.293 & 0.140 & 0.941 & 0.925 & 0.937 & 0.910 \\
% & \qwen     & 23.42 & 44.59 & 0.206 & 0.516 & 0.338 & 0.235 & 0.977 & 0.974 & 0.986 & 0.968 \\
% \bottomrule
% \end{tabular}

\begin{tabular}{llcccccccccccc}
\toprule
% \textbf{Dataset
& \multirow{2}{*}{\textbf{Model}} &
\multicolumn{2}{c}{\textbf{Overrated (\%)}} &
\multicolumn{2}{c}{\textbf{Underrated (\%)}} &
\multicolumn{2}{c}{\textbf{Mean Bias}} &
\multicolumn{2}{c}{\textbf{Cohen's $\kappa$}} &
\multicolumn{4}{c}{\textbf{Avg Confidence}} \\

\cmidrule(lr){3-4}
\cmidrule(lr){5-6}
\cmidrule(lr){7-8}
\cmidrule(lr){9-10}
\cmidrule(lr){11-14}

& &
\textbf{Bin.} & \textbf{Grad.} &
\textbf{Bin.} & \textbf{Grad.} &
\textbf{Bin.} & \textbf{Grad.} &
\textbf{Bin.} & \textbf{Grad.} &
\textbf{TP} & \textbf{FP} & \textbf{TN} & \textbf{FN} \\
\midrule

\multirow{4}{*}{\rotatebox[origin=c]{90}{\textbf{DL19}}}
& \gemma    & 37.81  & 61.84 & 3.74 & 8.36 & 0.341  & 0.736 & 0.246 & 0.107 & 0.987 & 0.989 & 0.989 & 0.977 \\
& \llama    & 18.82  & 49.72 & 11.09 & 12.17 & 0.077  & 0.564 & 0.304 & 0.163 & 0.902 & 0.881 & 0.958 & 0.914 \\
& \mistral  & 21.53  & 54.23 & 7.45 & 10.45 & 0.141  & 0.557 & 0.369 & 0.150 & 0.939 & 0.932 & 0.951 & 0.904 \\
& \qwen     & 20.62  & 47.02 & 6.10 & 10.3 & 0.145  & 0.459 & 0.421 & 0.215 & 0.977 & 0.976 & 0.991 & 0.986 \\
\midrule

\multirow{4}{*}{\rotatebox[origin=c]{90}{\textbf{DL20}}}
& \gemma    & 44.78 & 66.51 & 1.66 & 5.36 & 0.431 & 0.895 & 0.163 & 0.088 & 0.983 & 0.979 & 0.986 & 0.981 \\
& \llama    & 18.57 & 51.11 & 6.21 & 8.00 & 0.124 & 0.691 & 0.265 & 0.149 & 0.908 & 0.870 & 0.920 & 0.871 \\
& \mistral  & 22.32 & 55.08 & 4.35 & 7.47 & 0.180 & 0.646 & 0.293 & 0.140 & 0.941 & 0.925 & 0.937 & 0.910 \\
& \qwen     & 23.42 & 44.59 & 2.79 & 6.61 & 0.206 & 0.516 & 0.338 & 0.235 & 0.977 & 0.974 & 0.986 & 0.968 \\
\bottomrule
\end{tabular}

%% file: tables/pair_acc.tex
\begin{tabular}{llcccccccccc}
\toprule
& \multirow{3}{*}{\textbf{Model}} &
\multicolumn{4}{c}{\textbf{Pairwise Metrics (\%)}} &
\multicolumn{4}{c}{\textbf{Avg Confidence}} \\

\cmidrule(lr){3-6}
\cmidrule(lr){7-10}

& &
\textbf{\makecell{Acc.}} & \textbf{\makecell{Acc.\\(No Ties)}} & \textbf{\makecell{Tie\\Rate}} & \textbf{\makecell{Wrong\\Wins}} &
\textbf{Corr.} & \textbf{Incorr.} \\
\midrule

\multirow{4}{*}{\rotatebox[origin=c]{90}{\textbf{DL19}}}
& \gemma    & 50.35 & 89.47 & 43.72 & 5.92 & 0.979 & 0.963 \\
& \llama    & 57.59 & 84.65 & 31.95 & 10.44 & 0.981 & 0.978\\
& \mistral  & 56.50 & 87.77 & 35.61 & 7.87 & 0.978  & 0.961 \\
& \qwen     & 50.58 & 92.71 & 45.44 & 3.97 & 0.974 & 0.946 \\
\midrule

\multirow{4}{*}{\rotatebox[origin=c]{90}{\textbf{DL20}}}
& \gemma    & 53.78 & 89.52 & 39.92 & 6.29 & 0.989 & 0.970 \\
& \llama    & 60.01 & 85.46 & 23.78 & 10.20 & 0.982 & 0.981 \\
& \mistral  & 61.36 & 88.19 & 30.42 & 8.21 & 0.984 & 0.964 \\
& \qwen     & 54.42 & 93.08 & 41.52 & 4.04 & 0.982 & 0.962  \\
\bottomrule
\end{tabular}

% \begin{tabular}{cccccc}
% \toprule
% \textbf{Model} & \textbf{Data} & \textbf{Total Acc} & \textbf{Acc (No Ties)} & \textbf{Tie Rate} & \textbf{Wrong Wins} \\
% \midrule

% \multirow{2}{*}{\gemma} 
% &DL19 & 50.35\% &89.47\% & 43.72\% & 5.92\%  \\
% &DL20 &53.78\% & 89.52\% & 39.92\% & 6.29\%  \\
% \midrule

% \multirow{2}{*}{\mistral} 
% &DL19 & 56.50\% & 87.77\% & 35.61\%  & 7.87\%  \\
% &DL20 & 61.36\% & 88.19\% & 30.42\% & 8.21\% \\
% \midrule

% \multirow{2}{*}{\llama}  
% &DL19 & 57.59\%  & 84.65\% & 31.95\% & 10.44\% \\
% &DL20 & 60.01\%  & 85.46\% & 23.78\% & 10.20\%  \\
% \midrule

% \multirow{2}{*}{\qwen} 
% &DL19 & 50.58\% & 92.71\% & 45.44\% & 3.97\% \\
% &DL20 & 54.42\%  & 93.08\% & 41.52\% & 4.04\% \\

% \bottomrule
% \end{tabular}

%% file: 3_relevance_cues.tex
\section{Exploring Relevance Cues}
\begin{figure}[t]
    \centering
    \includegraphics[width=1\linewidth]{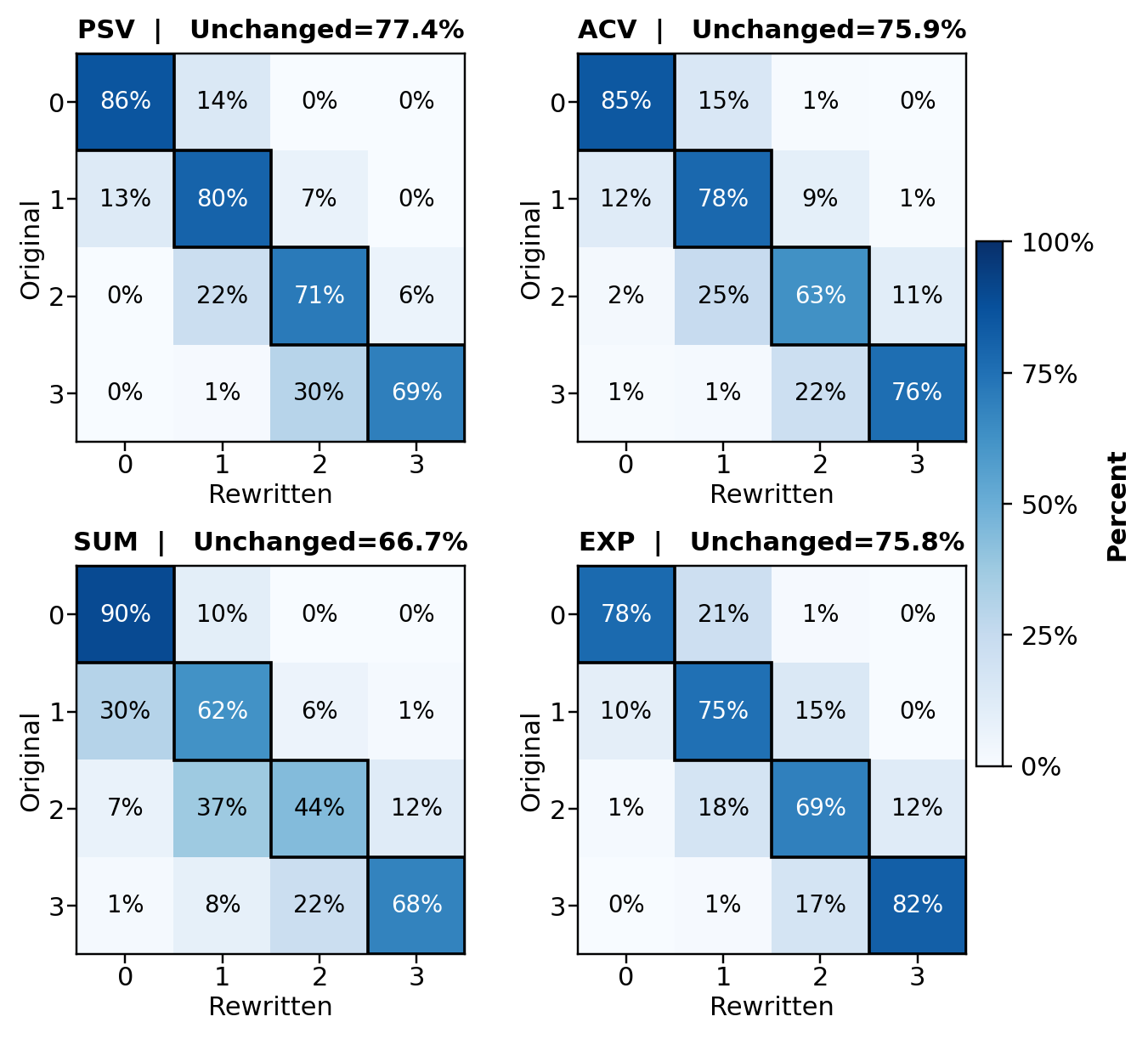}
    \vspace{-0.5cm}
    \caption{Label transition matrices for {\qwen} under different passage rewrites.
    Each heatmap shows normalized transitions from the LLM's original assigned relevance labels.}
    \label{fig:qwen-rewrite}
    \vspace{0.2cm}
\end{figure}
The overrating behaviors observed earlier raise a key question: do LLMs genuinely assess semantic relevance, or are their judgments disproportionately driven by surface-level lexical and structural cues? We evaluate robustness to semantically equivalent structural variations (syntactic and length changes), where label divergence indicates reliance on surface cues. We then introduce controlled, query-dependent sentence insertions into non-relevant passages to disentangle semantic relevance from lexical overlap.

% The overrating behaviors demonstrated in the previous set of experiments raise a key question: do LLMs genuinely assess semantic relevance, or are their judgments disproportionately driven by surface-level lexical and structural cues? We first examine robustness to semantically equivalent structural variations (i.e., syntactic and length variations), where score divergence indicates reliance on surface cues. We then introduce controlled, query-dependent sentence insertions into non-relevant passages to disentangle semantic relevance from lexical overlap.

% These overrating behaviors raise a broader question: do LLMs genuinely assess semantic relevance, or are their relevance decisions disproportionately influenced by surface-level lexical and structural cues? To investigate this question, we first examine robustness to semantically equivalent variations as an indicator of overrating. Divergent relevance scores assigned to meaning-preserving rewrites suggest reliance on structural cues (e.g., passage length or syntactic form) rather than true semantic relevance. We then construct three controlled query-related variants, each consisting of a single sentence, and insert them into irrelevant passages to further disentangle semantic relevance from lexical overlap.

% \subsection{Structural Variations with Preserved Semantics}
\subsection{Semantic-Preserving Structural Variations}
\label{semantic}
We evaluate judgment consistency across four meaning-preserving variants of each passage to quantify sensitivity to structural form.

% We analyze whether judgments remain consistent on four different variations of each passage. This approach quantifies surface sensitivity while exposing the model's intrinsic preferences for specific structural forms. 

\myparagraph{Syntactic Variation.}
LLMs may favor scientific writing styles~\cite{Faggioli:2023:Perspect}, where passive constructions are more common~\cite{fowler2015fowler}. To test whether such stylistic cues influence relevance judgments, we rewrite passages using passive (\passive) and active (\activ) voice, following expert recommendations favoring the active voice for clarity~\cite{lubis2024passive}. %Differences in assigned relevance scores indicate whether LLMs conflate a ``scientific tone'' with relevance.

\myparagraph{Length Variation.}
LLMs are known to prefer longer responses regardless of quality~\cite{zheng2023JudgingLLMasaJudgeMTBench,saito2023verbosity}, and input length shifts LLM label distributions in model-dependent ways~\cite{mohtadi2026effect}. We generate summarized (\summary) and expanded (\expansion) passage variants while %explicitly 
ensuring semantic equivalence. Comparing relevance scores across these variants reveals whether passage length is mistakenly treated as a relevance signal.

\myparagraph{Generating Structural Variations.}
Findings from~\citet{balog:2025:rankersjudgesassistants} indicate no general bias from LLM-judges to LLM-generated content, even within the same model family. We use {\gemini} to generate stylistic and length-based content variations. Unlike~\citet{balog:2025:rankersjudgesassistants}, who assume rewriting preserves relevance (validated by human assessors~\cite{dai:2024:neuralretrieversarea}), we perform an explicit quality-check by prompting {\gemini} to verify that all rewritten passages preserve the meaning of the original passage. Across all prompts, we retain only the passages identified as ``meaning preserved.''

\myparagraph{Sample Strategy.}
For each query, we independently sample (at maximum) 20 passages per predicted relevance label (0–3) for each LLM. For each passage, we generate four syntactic variants and assess label consistency across versions to analyze intrinsic model preferences. To avoid redundant rewriting from overlapping samples across LLMs, passages are pooled only at the rewriting stage, while each LLM is re-judged and evaluated solely on its own sampled query–rewritten passage pairs.
% We evaluate consistency, mean bias, score distribution shifts between original and rewritten passages, and Cohen's~$\kappa$.

\myparagraph{Results.}
% \textit{Rewritten Passages Judgment Results.}
Figure~\ref{fig:qwen-rewrite} shows the results as label transition matrices, capturing the change in judgment due to the four different passage re-writing treatments as compared to the original (unedited passage) LLM judgment.
Across all models, passive voice rewrites and active voice rewrites produce nearly identical effects on LLM judgments, indicating no systematic preference for one form over the other. Summarization consistently reduces relevance labels across models, whereas passage expansion does not consistently inflate relevance labels. Passage expansion systematically inflates relevance labels for originally low-relevance passages -- but can still degrade labels for highly relevant passages. These results indicate that even when semantic meaning is preserved, LLM-based relevance judgments exhibit a clear tendency to disfavor short passages.

% \textit{Syntactic Variation.} Large language models (LLMs) may be prone to regard scientific documents as more relevant~\cite{Faggioli:2023:Perspect}. Since passive constructions are more common in scientific writing~\cite{fowler2015fowler}, this stylistic preference may influence relevance judgments. However, expert guidelines recommend using the active voice rather than the passive voice to improve clarity in academic writing~\cite{lubis2024passive}. To isolate this effect, we rewrite passages into Passive Voice (Prompt-1) and Active Voice (Prompt-2). Comparing scores across these opposing styles reveals whether LLMs mistakenly the ``scientific tone'' of passive voice with higher relevance.

% \textit{Length Variation.}
% LLMs tend to prefer longer responses\cite{zheng2023JudgingLLMasaJudgeMTBench, saito2023verbosity}, even when increased length does not correspond to higher quality. We investigate the effect of passage length on LLM relevance judgments by generating summarized (Prompt-3) and expanded (Prompt-4) variants. Unlike prior length-based studies, we apply explicit quality control to ensure that both variants preserve the original semantic content. By comparing relevance scores assigned to summarized and expanded passages, we assess whether LLM judges conflate passage length with relevance signal.

\subsection{Controlled Lexical–Semantic Variations}
\label{subsubsec:query_varaint_injecttion}

\begin{figure}[t]
\centering
\begin{tcolorbox}[
    colback=gray!5, colframe=gray!40!black,
    fonttitle=\bfseries\small, fontupper=\small
]
\textbf{\textsf{QID:}} 156493 \\[0.4ex]
\textbf{\textsf{QUERY:}} \textit{do goldfish grow}
\vspace{2pt}
\begin{description}[leftmargin=3em, style=nextline,, itemsep=3pt]
  \item[\textsf{SEM}] These orange cyprinids continue to increase in size throughout their lifespan under suitable conditions.
  \item[\textsf{LEX}] The terms \textit{goldfish} and \textit{grow} are shown together as an example of word grouping in text data.
  \item[\textsf{QRY}] do goldfish grow.
\end{description}
\end{tcolorbox}
\vspace{-4pt}
\caption{\textsf{SEM}, \textsf{LEX}, and \textsf{QRY} variants for the query
\mbox{\textit{``do goldfish grow''}}. \textsf{SEM} preserves relevance
without surface query terms; \textsf{LEX} provides the query terms such that they
are not relevant to the information need; and \textsf{QRY} simply adds the query itself.}
\label{fig:goldfish-example}
\vspace{3pt}
\end{figure}

Prior work shows that query terms can cause LLMs to overrate \textit{non-relevant} passages~\cite{Alaofi:2024:LLMs-Can}. To disentangle lexical overlap from semantic relevance, we construct three controlled, query-dependent variants, including Semantic-only (\sem), Lexical-only (\lex), and Query-only (\qry), by inserting \textit{a single sentence} at the beginning of passages originally labeled non-relevant by both human assessors and LLMs, where injection has the strongest influence on relevance judgments~\cite{tamber2025IllusionsRelevanceUsing}. \Cref{fig:goldfish-example} illustrates these three variants for the example query ``do goldfish grow'', where each injected sentence is prepended to non-relevant passages to re-assess their relevance.

For the {\sem} variant, we prompt {\gemini} to generate a single sentence that directly answers the query while excluding all query terms after stopword removal. When exact synonyms are unavailable, dictionary-style paraphrases are used to preserve meaning without introducing lexical overlap. All generated sentences are manually verified for semantic correctness and constrained to minimal length to avoid confounding passage-length effects.

For the {\lex} variant, we construct a semantically non-relevant sentence that retains all query terms (excluding stopwords). To strictly control unigram overlap and eliminate unintended changes to the passage meaning, we use a fixed, content-neutral template:
``{\textit{The terms $w_1$, $w_2$, \ldots, $w_n$ are shown together as an example of word grouping in text data}}.'', where $w_k$ represent the $k$ individual query terms.

In the {\qry} variant, the inserted sentence is the original query string itself~\cite{Alaofi:2024:LLMs-Can}. This design aims to maximize full-query lexical overlap with the non-relevant passage without introducing additional semantic content beyond the query.

\begin{figure}[t]
    \centering
    \includegraphics[width=1\linewidth]{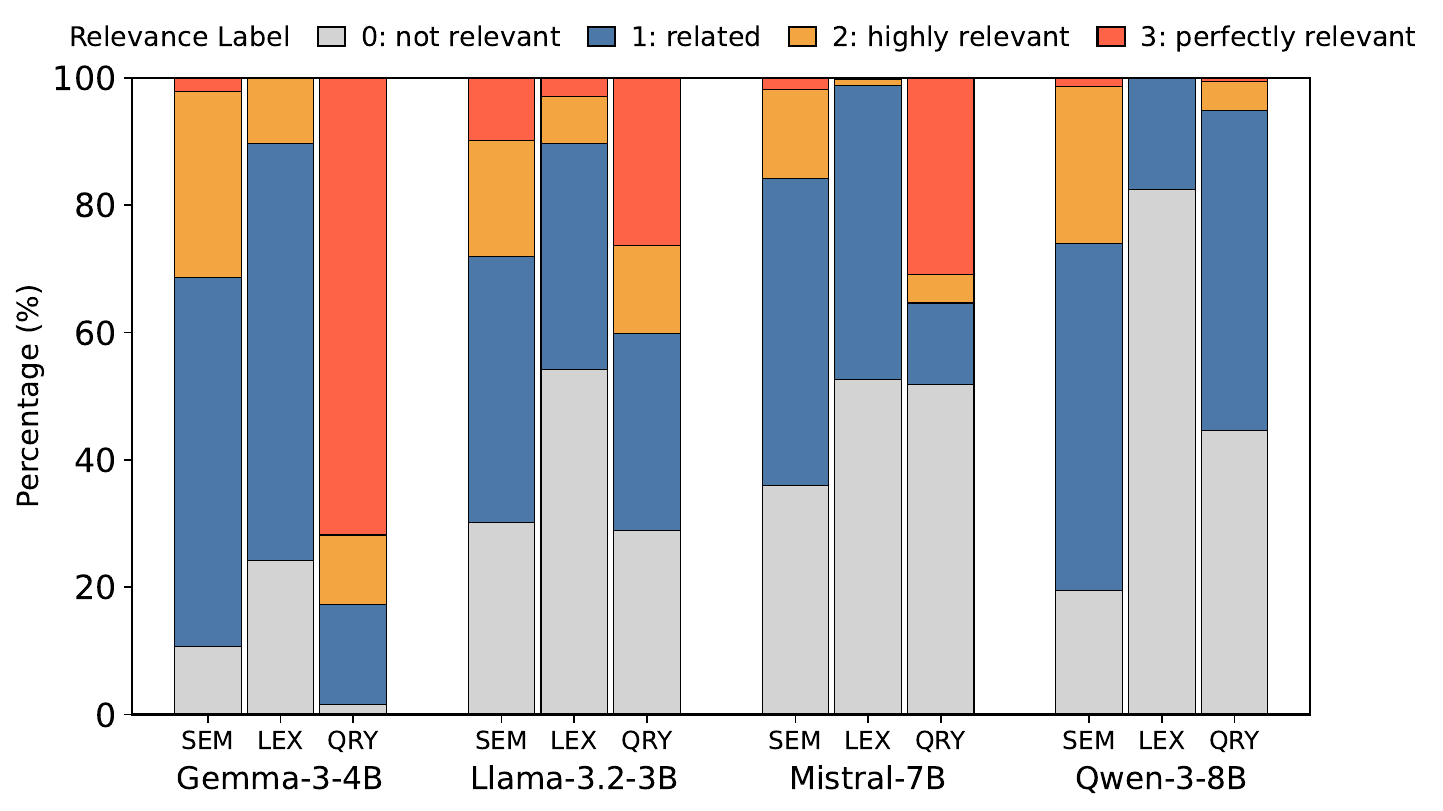}
    \vspace{-0.3cm}
    \caption{Predicted relevance-label distributions on non-relevant passages after three query-dependent insertions.}
    \label{fig:model-abc-dist}
    \vspace{6pt}
\end{figure}

\myparagraph{Results.}
% \textit{Controlled Lexical–Semantic Variations Results.}
Figure~\ref{fig:model-abc-dist} reports the label distributions after inserting each variant. Under the {\sem} condition, the majority of the models assign 1--2 relevance labels, with \textit{perfectly relevant} (label 3) rarely assigned, indicating systematic \textit{under-recognition of semantic relevance} in the absence of lexical cues. Under {\lex} condition, unigram-level lexical overlap alone is sufficient to elevate relevance labels despite the absence of semantic alignment. This reveals a strong sensitivity to term-level lexical cues. While some models (e.g., Gemma) assign labels comparable to \sem, others (e.g., Qwen) remain largely at label 0, suggesting that query terms are insufficient to override semantic non-relevance for stronger models. The {\qry} condition yields severe overrating: inserting only query string frequently triggers ~\emph{perfectly relevant} labels for non-relevant passages which confirms prior findings~\cite{Alaofi:2024:LLMs-Can}. Gemma assigns \emph{perfectly relevant} to the majority of non-relevant passages once the full query is inserted, while others (Mistral, Qwen) remain relatively robust. This demonstrates that full-query lexical anchoring can override semantic judgment in most cases. Overall, even with semantic understanding, our experiments show that LLM relevance assessment is disproportionately influenced by lexical cues, confirming strong lexical anchoring effects across models.

%% file: 4_conclusion.tex
%\vspace{-12pt}
\section{Conclusion}
\vspace{2pt}
In this study, we investigated label overrating and model confidence in LLM-based relevance judgments across multiple open-weight LLMs and evaluation paradigms.
Across all evaluation settings, we showed that overrating is a system-wide bias, particularly in graded and pairwise evaluations. Overrating is less pronounced in binary relevance judgments, suggesting that LLMs may be better suited for coarse-grained tasks that focus on distinguishing relevant from non-relevant passages. 
Through experiments on semantically equivalent passage variations, we demonstrate that LLM judges are sensitive to passage length even when semantic content is preserved, while showing comparatively limited sensitivity to syntactic form. Controlled query-related variant experiments indicate that, despite some semantic understanding, LLMs remain strongly influenced by lexical cues, especially direct query-string overlap. These findings underscore the need for diagnostic evaluation frameworks that explicitly probe such biases when applying LLMs for relevance assessment.

%% file: 5_future_works.tex
\section{Future Works}
The experimental setup relies only on TREC DL datasets, which primarily consist of short queries paired with relatively short passages. Future work will extend our analysis to investigate how LLM-based relevance judges behave in collections, such as TREC-8~\cite{hawking1999overview} and TREC Robust04~\cite{voorhees2005overview}, contain substantially longer, multi-topic documents. Another concern is our current study adopts the UMBRELA prompt as a representative instantiation of LLM-based relevance judges. Prior work demonstrates that variations in prompt structure can substantially influence LLM judgment behaviors. Extending beyond a single prompt is therefore essential to establish the robustness of our findings. Future work will construct analysis on multiple prompts spanning reasoning strategies, prompt structures, and persona. These extensions will strengthen the generalizability of our findings and position the proposed analysis as a framework for studying LLM judges, rather than a result tied to a single prompt instantiation. 

We acknowledge the concern that modern LLMs may have been exposed to MS MARCO~\cite{nguyen2017ms} or TREC Deep Learning~\cite{craswell2020overview,craswell2021overview} qrels during pre-training~\cite{Thomas:2024:Large-La,Faggioli:2023:Perspect}, which could inflate apparent agreement with ground-truth labels through memorization rather than genuine judgment capability. Our experimental design partially mitigate this risk. Specifically, the summarization and expansion rewriting generates passages that are unlikely to have appeared in pre-training, and the \texttt{SEM}, \texttt{LEX}, and \texttt{QRY} variants are synthetically constructed rather than directly drawn from the original collection. Hence, the bias patterns we report in this work cannot be fully explained by memorization of specific query-passage pairs. However, we cannot fully rule out that LLMs retain topical or stylistic priors from MS~MARCO that bias their judgments in ways not captured by our controlled perturbations. A rigorous contamination audit remains an important future work.